\PassOptionsToPackage{table,usenames,dvipsnames}{xcolor}
\documentclass[10pt,a4paper,hidelinks]{article}
\newcommand{\mytitle}{Building Fast Fuzzers}

\usepackage{fancyhdr}
\usepackage{pdfpages}
\usepackage[procnames]{listings}
\usepackage{graphicx}

\fancypagestyle{CISPA}{

\fancyhead[C]{\includegraphics[width=4cm]{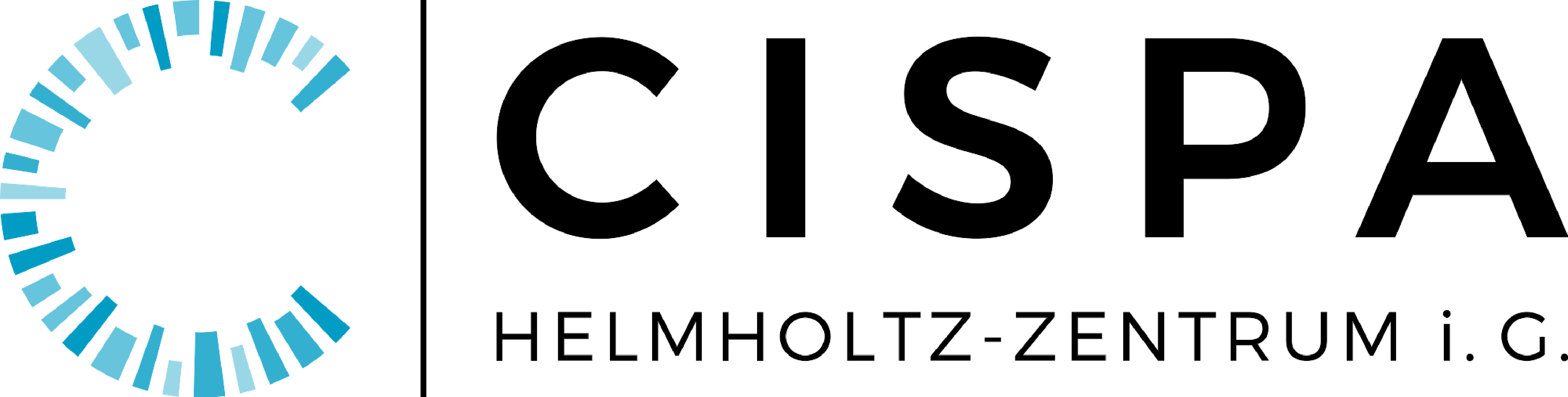}}
}
\usepackage[utf8]{inputenc}
\usepackage{csquotes}
\usepackage[backend=biber]{biblatex}
\usepackage{etoolbox}
\newif\ifdraft\draftfalse
\newif\iflong\longtrue

\usepackage[T1]{fontenc}
\usepackage[english]{babel}
\usepackage[nohead,includefoot,margin=2cm]{geometry}
\usepackage[compact,raggedright,small]{titlesec}
\usepackage[affil-it]{authblk}
\usepackage[runin]{abstract}
\usepackage{multicol}
\usepackage{titling}
\usepackage{montserrat}
\usepackage{setspace}

\usepackage{booktabs}   
\usepackage{multirow}
\usepackage{subcaption} 
\usepackage{caption}

\usepackage{hyperref}
\hypersetup{colorlinks=true,citecolor=blue,linkcolor=blue,filecolor=blue,urlcolor=blue,pdftitle={\mytitle}}
\usepackage{amsmath,amssymb,amsfonts}
\usepackage[nameinlink,noabbrev]{cleveref}  
\usepackage[inline]{enumitem}
\usepackage{microtype} 
\usepackage{alltt}
\usepackage{float}
\usepackage{stfloats}

\usepackage[noend]{algpseudocode}
\usepackage{color}
\usepackage{relsize} 

\usepackage{comment}
\usepackage{algorithm}
\usepackage{textcomp}
\usepackage{xspace}
\usepackage{syntax}
\usepackage{qtree}
\ifdraft
\usepackage{todonotes}
\else
\usepackage[disable]{todonotes}
\fi
\usepackage{tabularx}

\definecolor{codegreen}{rgb}{0,0.6,0}
\definecolor{codegray}{rgb}{0.5,0.5,0.5}
\definecolor{codepurple}{rgb}{0.58,0,0.82}
\definecolor{backcolour}{rgb}{0.95,0.95,0.92}

\lstdefinestyle{mystyle}{
    commentstyle=\color{codegreen},
    keywordstyle=\color{magenta},
    numberstyle=\tiny\color{codegray},
    stringstyle=\color{codepurple},
    basicstyle=\footnotesize,
    breakatwhitespace=false,
    breaklines=true,
    captionpos=b,
    keepspaces=true,
    numbers=left,
    numbersep=5pt,
    showspaces=false,
    showstringspaces=false,
    showtabs=false,
    tabsize=2
}

\algblockdefx[switch]{Switch}{EndSwitch}%
[1]{\textbf{switch} #1}%
{}%
\algloopdefx{Case}[1]{\textbf{case} #1:}%
\algdef{SE}[DOWHILE]{Do}{doWhile}{\algorithmicdo}[1]{\algorithmicwhile\ #1}%

\def\|#1|{\textit{#1}}
\def\<#1>{\texttt{#1}}

\lstdefinestyle{grammar}
{
    basicstyle=\small\ttfamily,
    keywordstyle=\bfseries,
    numberblanklines=false,
    language=ebnf,
    tabsize=2,
    keywordstyle=\color{blue},
    identifierstyle=\color{red},
}

\definecolor{eclipseBlue}{RGB}{42,0.0,255}
\definecolor{eclipseGreen}{RGB}{63,127,95}
\definecolor{eclipsePurple}{RGB}{127,0,85}

\lstdefinestyle{python}
{
    basicstyle=\footnotesize\ttfamily,
    numberblanklines=false,
    language=python,
    tabsize=2,
    commentstyle=\color{eclipseGreen},
    keywordstyle=\bfseries\color{eclipsePurple},
    stringstyle=\color{eclipseBlue},
    procnamestyle=\bfseries\color{black},
    procnamekeys={def},
    identifierstyle=
}

\def\BibTeX{{\rm B\kern-.05em{\sc i\kern-.025em b}\kern-.08em
    T\kern-.1667em\lower.7ex\hbox{E}\kern-.125emX}}

\usepackage{framed}

\title{\mytitle}    
\date{\small (Dated \today)}
\author{Rahul Gopinath}
\author{Andreas Zeller}
\affil{\{rahul.gopinath, zeller\}@cispa.saarland \\
CISPA - Helmholtz Center for Information Security, Saarbr\"ucken, Germany}

\newcommand\BackgroundPic{
    \put(0,0){
    \parbox[b][\paperheight]{\paperwidth}{%
    \vfill
    \centering
    \includegraphics[width=\paperwidth,height=\paperheight]{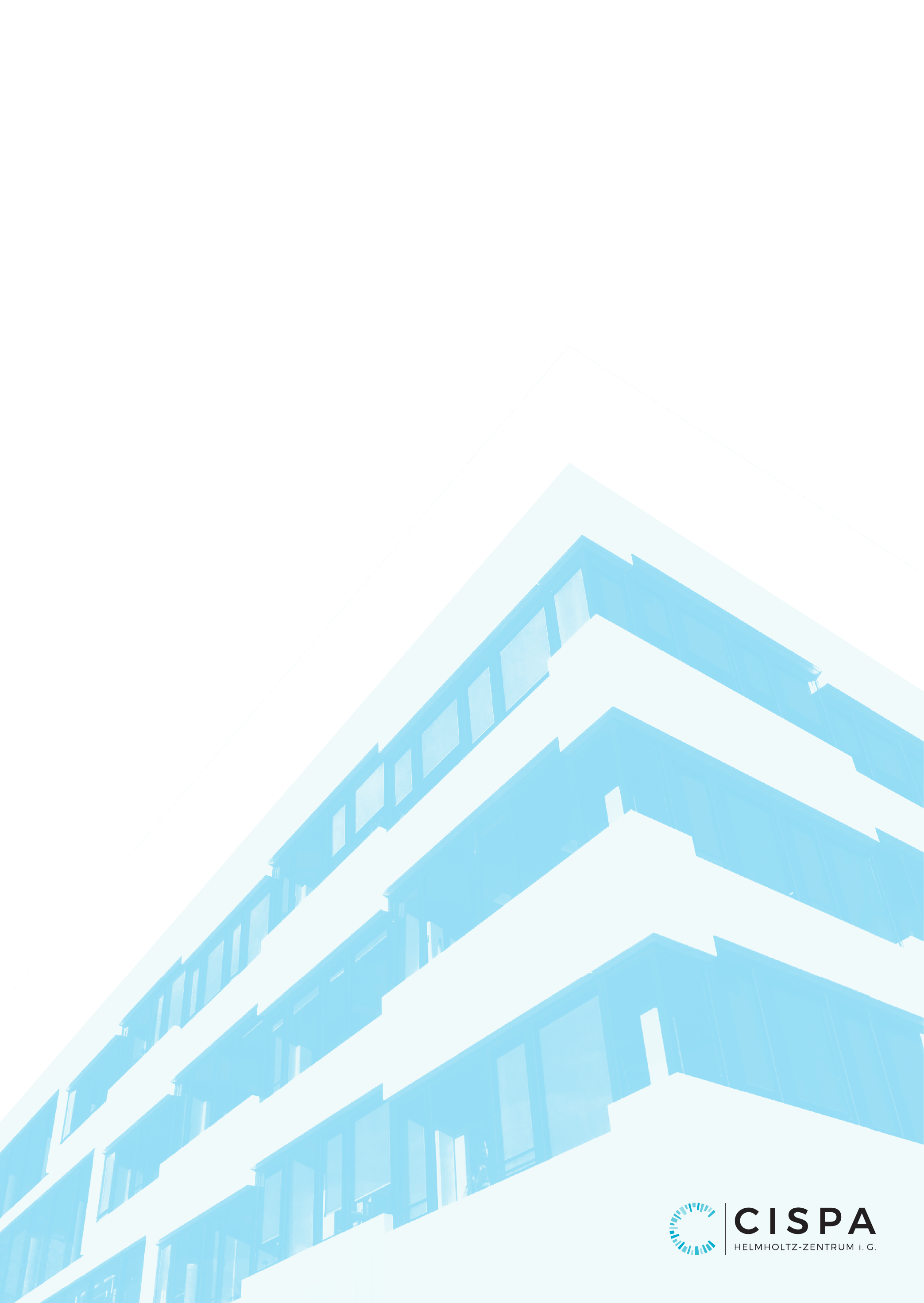}
    \vfill
}}}

\usepackage{utopia}
\usepackage{times}

\newcommand{\fone}{\textit{F1}\xspace}
\newcommand{\expr}{\textit{Expr}\xspace}

\newcommand{\dd}{\textit{dev-random}\xspace}
\newcommand{\nonterminal}{\textit{nonterminal}\xspace}
\newcommand{\terminal}{\textit{terminal}\xspace}

\newcommand{\warmuploop}{ten~\xspace}         
\newcommand{\numloops}{ten~\xspace}         

\newcommand{\XJSONcompiledPyThroughput}{365.45\xspace}

\newcommand{\XJSONsupercompiledPyThroughput}{573.76\xspace}

\newcommand{\XJSONcompiledCThroughput}{13,006.68\xspace}

\newcommand{\XJSONoptimizedMMAPCThroughput}{41,273.09\xspace}
\newcommand{\XJSONoptimizedCThroughput}{\XJSONoptimizedMMAPCThroughput}

\newcommand{\XJSONDirectThreadedThroughput}{40,750.50\xspace}

\newcommand\hmm[1]{\ifnum\ifhmode\spacefactor\else2000\fi>1000 \uppercase{#1}\else#1\fi}
\newcommand{\grammarinator}{\hmm{g}rammarinator\xspace}
\newcommand{\dharma}{\hmm{d}harma\xspace}
\newcommand{\gramfuzz}{\hmm{g}ramfuzz\xspace}

\usepackage{xspace}

\begin{document}
\AddToShipoutPicture*{\BackgroundPic}

\makeatletter
\renewcommand{\Authfont}{\normalsize\sffamily\bfseries}
\renewcommand{\Affilfont}{\normalsize\sffamily\mdseries}
\begin{titlepage}
\newcommand{\HRule}{\rule{\linewidth}{0.1mm}}
\centering
  \textsc{\LARGE {\fontfamily{Montserrat-TOsF}\selectfont CISPA Helmholtz-Zentrum i.G.}}\\[1.5cm]

  \vspace{2.4 cm}
  \HRule \\[0.2cm]
  {\huge\sffamily\bfseries \@title\par}
  \vspace{0.2cm}
  \HRule \\[1.5cm]

  {\sffamily \@author\par}
\vfill

\end{titlepage}

\makeatother
\setlength{\affilsep}{0.1em}
\addto{\Affilfont}{\small}
\renewcommand{\Authfont}{\normalsize}
\renewcommand{\Affilfont}{\normalsize}

\pretitle{\begin{center}\large\bfseries}
\posttitle{\end{center}}
\maketitle
\thispagestyle{CISPA}

\begin{abstract}

Fuzzing is one of the key techniques for evaluating the robustness of programs
against attacks.  Fuzzing has to be \emph{effective} in producing inputs that cover functionality and find vulnerabilities.  But it also has to be \emph{efficient} in producing such inputs quickly.  Random fuzzers are very efficient, as they can quickly generate random inputs; but they are not very effective, as the large majority of inputs generated is syntactically invalid.  Grammar-based fuzzers make use of a grammar (or another model for the input language) to produce syntactically correct inputs, and thus can quickly cover input space and associated functionality.  Existing grammar-based fuzzers are surprisingly inefficient, though: Even the fastest grammar fuzzer \dharma still produces inputs about
\emph{a thousand times slower} than the fastest random fuzzer.
So far, one can have an effective or an efficient fuzzer, but not both.

In this paper, we describe how to build fast grammar fuzzers from the ground up, treating the problem of fuzzing from a programming language implementation perspective.  Starting with a Python textbook approach, we adopt and adapt optimization techniques from functional programming and virtual machine implementation techniques together with other novel domain-specific optimizations in a step-by-step fashion.  In our \fone prototype fuzzer, these \emph{improve production speed by a factor of 100--300 over the fastest grammar fuzzer} \dharma.  As \fone is even 5--8 times faster than a lexical random fuzzer, we can find bugs faster and test with much larger valid inputs than previously possible.





%
%
%
%
%

\end{abstract}
\begin{multicols}{2}

\section{Introduction}
\label{sec:introduction}

Fuzzing is a popular technique for evaluating the robustness of programs
against attacks. The effectiveness of fuzzing comes from fast production and
evaluation of inputs and low knowledge requirements about the program or its
behavior---we only need to detect program crashes. These 
properties make fuzzing an attractive tool for security professionals.

To be effective, a fuzzer needs to sufficiently cover the variety of possible
inputs, and it should produce inputs that can reach deep code paths. To reach
deep code paths, the fuzzer needs to produce inputs that can get past the input
parser---i.e., inputs that conform to the \emph{input language} of the program
under test.

Hence, the effectiveness of fuzzing can be improved by incorporating knowledge
about the \emph{input language} of the program under test to the fuzzer. As
such languages are typically described by formal grammars, fuzzers that
incorporate language knowledge are called \emph{grammar fuzzers}. The inputs
produced by grammar fuzzers are superior to pure random fuzzers because
they easily pass through input validators, and hence a larger number of inputs can exercise deeper logic in the program.  Today, a large number of tools exist~\cite{hodovan2018grammarinator,wang2017skyfire,wang2019superion,aschermann2019nautilus,guo2013gramfuzz,devarajan2007unraveling,veggalam2016ifuzzer} that all provide grammar-based fuzzing.  These tools are also effective in their results: The LANGFUZZ grammar fuzzer~\cite{holler2012fuzzing} for instance, has uncovered more than 2,600 vulnerabilities in the JavaScript interpreters of Firefox and Chrome.

\begin{figure}[H]
  \begin{grammar}
    <START> ::= <expr>

    <expr>  ::= <term>
    \alt <term> `+' <expr>
    \alt <term> `-' <expr>

    <term>  ::= <factor>
    \alt <factor> `*' <term>
    \alt <factor> `/' <term>

    <factor>  ::= <integer>
    \alt <integer> `.' <integer>
    \alt `+' <factor>
    \alt `-' <factor>
    \alt `(' <expr> `)'

    <integer> ::= <digit> <integer>
    \alt <digit>

    <digit> ::= `0' | `1' | `2' | `3' | `4' | `5' | `6' | `7' | `8' | `9'
  \end{grammar}
\caption{A grammar for arithmetic expressions}
  \label{fig:contextfreeexpr}
\end{figure}
\begin{figure}[H]
\lstset{
 morekeywords={start,expr,term,factor,integer,digit},
 moredelim={[is][stringstyle]{'}{'}},
}
\begin{lstlisting}[style=python, numbers=left, numberstyle=\tiny]
expr_grammar = {
    '<start>':   (['<expr>']),
    '<expr>':    (['<term>', "+", '<expr>'],
                  ['<term>', "-", '<expr>'],
                  ['<term>']),
    '<term>':    (['<factor>', "*", '<term>'],
                  ['<factor>', "/", '<term>'],
                  ['<factor>']),
    '<factor>':  (["+", '<factor>'],
                  ["-", '<factor>'],
                  ["(", '<expr>', ")"],
                  ['<integer>', ".", '<integer>'],
                  ['<integer>']),
    '<integer>': (['<digit>', '<integer>'],
                  ['<digit>']),
    '<digit>':   (["0"], ["1"], ["2"], ["3"], ["4"].
                  ["5"], ["6"], ["7"], ["8"], ["9"])
}
\end{lstlisting}
\caption{The grammar from \Cref{fig:contextfreeexpr} as a Python \emph{dict}}

\label{fig:contextfreeexprpy}
\end{figure}

Grammar fuzzers have two downsides, though. The first problem is that an
input grammar has to exist in the first place before one can use them for
fuzzing. Programmers may choose not to use a formal grammar for describing the
input, preferring to use ad hoc means of specifying the input structure. Even
when such a grammar is available, the grammar may be incomplete or obsolete,
and fuzzers relying on that grammar can develop blind spots.  Recently, however, a number of approaches have been proposed
to infer both regular languages~\cite{walkinshaw2008evaluation}
as well as context-free grammars either by grammar induction from
samples~\cite{bastani2017synthesizing} or by dynamic grammar inference from program code~\cite{hoschele2017active}.  While these approaches require valid sample inputs to learn from, recent work by Mathis et al.~\cite{mathis2019parser} and Blazytko et al.~\cite{blazytko2019usenix} suggests that it is possible to automatically generate inputs that cover all language features and thus make good samples for grammar induction---even for complex languages such as JavaScript.  It is thus reasonable to assume that the effort for specifying grammars might be very reduced in the future.

\begin{figure}[H]
\begin{lstlisting}[style=python, numbers=left, numberstyle=\tiny]
def gen_key(grammar, key):
   if key not in grammar: return key
   return gen_alt(grammar, random.choice(grammar[key]))

def gen_alt(grammar, alt):
    # Concatenate expansions of all elements in alt
    return ''.join(gen_key(grammar, t) for t in alt)

gen_key(expr_grammar, '<start>')
\end{lstlisting}
\caption{A simple grammar based fuzzer in Python that uses the grammar from \Cref{fig:contextfreeexprpy}}
\label{fig:simplefuzzer}
\end{figure}


The second problem with grammar fuzzing, though, is that it is \emph{slow}---at
least in the implementations we see today.  In principle, fuzzing with a grammar
is not hard. Given a context-free grammar such as in~\Cref{fig:contextfreeexpr},
we first convert it to a native data structure as given in~\Cref{fig:contextfreeexprpy}.
We then start with the \nonterminal representing the starting point
\texttt{<start>}. Each \nonterminal
corresponds to possibly multiple alternative rules of expansion.
The fuzz result corresponds to one of the rules of expansion for
\texttt{<start>}, choosing expansions randomly.

An expansion is a sequence of tokens. For each token, if the token
is a \terminal symbol (a string with no further expansion), then the string
goes unmodified to its corresponding location in the final string. If it is a
\nonterminal, we start expansion again until the expansion results in \terminal
nodes. A Python program implementing this approach is given
in~\Cref{fig:simplefuzzer}.

Unfortunately, this simple approach has a problem when dealing
with highly recursive grammars. As the fuzzer uses the recursion as a method to
explore nested structures, it can deplete the stack quite fast. This can be
fixed by limiting the alternatives explored to the lowest cost alternatives.
We show how this can be done in \Cref{sec:depth}.

However, once the problem of stack exhaustion is fixed, we have another problem
While this production approach is not na\"\i{}ve, it is not exactly fast either.
Using a simple \expr grammar from the ``Fuzzing Book'' textbook chapter on
Grammars~\cite{fuzzingbook2019:Grammars}, 
it provides a throughput of 103.82 kilobytes per second.\footnote{In this paper, 1 ``kilobyte'' = 1,024 bytes = 1 KiB, and 1 ``megabyte'' = 1,024 $\times$ 1,024 bytes = 1 MiB.} If one wants \emph{long}
inputs of, say, ten megabytes to stress test a program for buffer and
stack overflows, one would thus have to wait for a minute to produce one single input.
Now, compare this to a pure random fuzzer, say using
\texttt{dd if=/dev/urandom} directly as input to a program (that we call the
\dd fuzzer). Using \dd achieves a speed of 23~MiB per
second, which is over \emph{a hundred times faster} than even the fastest
grammar-based fuzzer \dharma, which produces 174.12 KiB/s on \emph{Expr}.




In this paper, we \emph{show how to build fast grammar fuzzers from the ground up,} treating the problem of fuzzing from a programming language implementation
perspective.  Starting from the above Python code, we apply and develop a number of techniques to turn it into an extremely fast fuzzer.  On a CSS grammar, our \fone prototype yields a final throughput of 80,722 kilobytes per second of valid inputs on a single core.  This is \emph{333 times faster} than \dharma, the fastest grammar fuzzer to date (which only produces 242 KiB/s for CSS), and even three times as fast as the \dd fuzzer.  These results make our
\fone prototype the fastest fuzzer in the world that produces valid inputs---by a margin.

A fast fuzzer like \fone not only can be used to generally save CPU cycles for
fuzzing, but also to produce large inputs for areas in which high throughput is required, including CPUs~\cite{martignoni2009testing},
FPGAs~\cite{laeufer2018rfuzz}, Emulators~\cite{domas2017breaking},
and IOT devices~\cite{zheng2019firm}, all of which allow high speed
interactions.  It also allows for stress testing hardware encoders and decoders
such as video encoders that require syntactically valid, but
rarely seen inputs can profit from a fast grammar fuzzer.
Side channel attacks on hardware implementing encryption and decryption
may require that the certificates use an envelope with valid values,
with the encrypted values as one of the contained values, where again the
\fone fuzzer can help. Finally, many machines use a hardware implemented
TCP/IP stack, which require structured input. Given that the networking
stack can accept high throughput traffic, \fone can be of use fuzzing
these devices. In all these settings, speed becomes a priority.

The remainder of this paper is organized as follows.  After discussing related work (\Cref{sec:background}) and introducing our
evaluation setup (\Cref{sec:evaluation}), we first discuss methods of limiting expansion depth (\Cref{sec:depth}).  We then make the following contributions:


\begin{description}
\item[Grammar Compilers.] We show how to \emph{compile} the grammar to \emph{program code,} which,
instead of interpreting a grammar structure, directly acts as a producer (\Cref{sec:compiling}).  A compiled producer is much faster than a grammar interpreter.

\item[Compiling to Native Code.]  We can further speed up production by directly compiling to native code, e.g. producing C code which is then compiled to native code (\Cref{sec:c}).  Compared to languages like Python (in which most grammar fuzzers are written), this again yields significant speedups.

\item[Supercompilation.] We introduce the novel notion of
\emph{supercompiling} the grammar, inlining production steps to a maximum (\Cref{sec:supercompiling}).  This results in fewer jumps and again a faster fuzzer.

\item[System Optimizations.] We explore and apply a number of \emph{system optimizations} (\Cref{sec:system}) that add to the efficiency of producers on practical systems, notably high-speed random number generation and quick file output.

\item[Production Machines.] We introduce the notion of a fuzzer as a \emph{virtual machine interpreter} (\Cref{sec:vm}) that interprets the random stream as bitcode, and explore the various alternatives in efficiently implementing virtual machines.

\item[The Fastest Fuzzer.]  In a detailed evaluation (\Cref{sec:comparison}) with multiple grammars and settings, we compare the performance of our \fone prototype implementation against state-of-the-art grammar fuzzers, and find it is 200--300 times faster than \dharma, the fastest grammar fuzzer to date.
\end{description}


After discussing related work (\Cref{sec:related}), \Cref{sec:limitations} discusses current limitations and future work.  Finally, \Cref{sec:conclusion} closes with conclusion and consequences.  All of the source code and data referred to in this paper is available as self-contained Jupyter Notebooks, allowing to run the \fone prototype and fully replicate and extend all experiments.

%

\section{Background}
\label{sec:background}

\subsection{Fuzzing and Test Generation}

Fuzzing is a simple but highly effective technique for finding vulnerabilities in
programs. The basic idea of a fuzzer is to quickly generate strings and evaluate these
strings as input to the program under fuzzing. If any of these inputs trigger a
program crash or other surprising behavior, or the program execution falls afoul
of sanity checks, it is an indication of a possible vulnerability that may be
exploited~\cite{fuzzingbook2019:Fuzzer}.

Fuzzers, and testing techniques in general, are traditionally classified into
\emph{whitebox} and \emph{blackbox} (and sometimes \emph{greybox})
techniques~\cite{manes2018fuzzing}.
Whitebox techniques assume availability of source code, and often use program
analysis tools to enhance the effectiveness of fuzzing, and example of which
is KLEE~\cite{cadar2008klee} which uses symbolic execution of source code for
generating inputs. Another is AFL~\cite{nossum2016filesystem}, which makes
use of specially instrumented binaries for tracking coverage changes.
Blackbox techniques on the other hand, do not require the
availability of source code. The greybox fuzzers assumes
the availability of at least the binary under fuzzing, and often work by instrumenting
the binary for recovering runtime information. An example of such a fuzzer is angr~\cite{shoshitaishvili2016sok} which can symbolically or concolically execute
a binary program to produce fuzzing inputs. The main problem with whitebox fuzzers
is the requirement of source code (and its effective utilization). The problem is
that time spent on analysis of source code can reduce the time spent on generating
and executing inputs.

The black box fuzzers (and the input generation part of the whitebox and greybox
fuzzers) are traditionally classified as \emph{mutational fuzzers} and generational fuzzers.

In mutational fuzzing, a corpus of seed inputs are used as the starting
point, which are then mutated using well defined operations to generate newer
inputs that can be evaluated~\cite{fuzzingbook2019:MutationFuzzer}. AFL,
libFuzzer~\cite{serebryany2016continuous} and Radamsa~\cite{pietikainen2011security}
are some of the well known mutational
fuzzers. Mutational fuzzers require relatively little knowledge of the input language
and relies on the given seed corpus to produce sufficiently valid
strings to reach deep code paths. One of the problems of mutational fuzzers is that
they are limited by the seed corpora. That is, these fuzzers are inefficient when
it comes to generating inputs that are markedly different from the ones in the
seed corpora. Unfortunately, these kinds of \emph{unexpected} inputs often have
higher chance of triggering bugs than the expected kind. This limits their utility.
A second problem is that due to the fuzzers ignorance of the input format, the
mutations introduced frequently fall afoul of the input validator often in trivial
ways. Hence, only a small portion of the mutated inputs reach the internal code paths
where the bugs hide. That is, if one is fuzzing an application that accepts a string
in an XML format, one might have more success in fuzzing the main application itself
rather than the XML parser itself which is likely to be well tested. Hence, it is
of interest to the fuzzer to generate valid XML fragments in the first place.

Generational fuzzers on the other hand, relies on some model of the input required
by the program to generate valid inputs. The model may be fixed as in the case of
Csmith~\cite{yang2011finding} which generates valid C programs and JSFunFuzz~\cite{ruderman2007introducing} which targets Javascript. Fuzzers such as Gramfuzz\cite{guo2013gramfuzz}, Grammarinator~\cite{hodovan2018grammarinator}, Dharma~\cite{mozilla2019dharma}, Domato~\cite{gratric2019domato},
and CSS Fuzz~\cite{ruderman2007introducing} allow the user to specify the
input format as a context-free grammar. For those contexts where a finite
state automata is sufficient, some fuzzers~\cite{cui2014novel,wang2017automatic} allow
an FSM as the input model. Fuzzers that allow context sensitive constraints on inputs
are also available~\cite{dewey2014language}.

\subsection{Context-Free Grammars}

A context-free grammar is a formal grammar that specifies how a set of strings
can be generated. A formal grammar is a set of rules (called production rules) for
rewriting a sequence of symbols. A production describes how a given symbol (called
a \nonterminal) should be replaced. If a symbol is not associated with a production
it is called a \terminal, and represents itself in the generated output. The
rewriting starts in the symbol representing the starting point, called the
\texttt{start} symbol. In the \Cref{fig:contextfreeexpr}, the start symbol is
\texttt{<start>}, and the production corresponding to it is \texttt{<expr>}.
Similarly, the \nonterminal symbol \texttt{<expr>} has three production rules:
\begin{lstlisting}[style=python, numbers=left, numberstyle=\tiny]
'<expr>': (['<term>', "+", '<expr>'],
           ['<term>', "-", '<expr>'],
           ['<term>']),
\end{lstlisting}
When fuzzing, one of these production rules is chosen stochastically for rewriting
\texttt{<expr>}. The first rule specifies that \texttt{<expr>} is rewritten as
a sequence \texttt{<term>+<expr>}, where \texttt{<term>} is again another
\nonterminal symbol while \texttt{+} is a \terminal symbol that is represented
by itself in the output.

The \nonterminal symbol \texttt{<term>} gets expanded to a string containing \texttt{<factor>} just like \texttt{<expr>} was expanded into a string containing \texttt{term}.  \texttt{<factor>} has five
production rules specifying how it may be expanded.
\begin{lstlisting}[style=python, numbers=left, numberstyle=\tiny]
'<factor>': (["+", '<factor>'],
             ["-", '<factor>'],
             ["(", '<expr>', ")"],
             ['<integer>', ".", '<integer>'],
             ['<integer>']),
\end{lstlisting}
If we assume that the production rule \texttt{<integer>} was chosen, then we
get to choose from the expansions of \texttt{<integer>} given by:
\begin{lstlisting}[style=python, numbers=left, numberstyle=\tiny]
'<integer>': (['<digit>', '<integer>'],
              ['<digit>']),
\end{lstlisting}
If we assume that the second production rule was chosen next, it contains a
single \nonterminal symbol \emph{<digit>}.
The \texttt{<digit>} has ten production rules, each of which has a single
\nonterminal symbol.
\begin{lstlisting}[style=python, numbers=left, numberstyle=\tiny]
    '<digit>':   (["0"], ["1"], ["2"], ["3"], ["4"].
                  ["5"], ["6"], ["7"], ["8"], ["9"])
\end{lstlisting}
If say \texttt{5} was chosen as the rule, then the first \texttt{<factor>}
would be replaced by \texttt{5}, giving the expression
\texttt{5*<factor>}. Similarly, the second \texttt{<factor>}
may also be replaced by say the fourth production rule:

\begin{center}
\texttt{factor} $\rightarrow$ \texttt{<integer>.<integer>}
\end{center}

\noindent
This gives the expression \texttt{5*<integer>.<integer>}.
Starting with unexpanded symbols on the left, assuming the second expansion
for \texttt{integer} was chosen, we have \texttt{digit}. Say \texttt{digit}
expanded to \texttt{3}, the above expression is transformed to:

\begin{center}
$\rightarrow$ \texttt{5*3.<integer>}
\end{center}

\noindent
Going through similar expansions for the last \texttt{integer} again, we get:

\begin{center}
$\rightarrow$ \texttt{5*3.8}
\end{center}

\noindent
which is the final expression.

Context-free grammars are one of the common ways to specify file formats, or as
the first level (parser) format for most programming languages.
The ability for patterns to be named, and reused makes it easier to use than
regular expressions, while the context-free aspect makes it easy to write a
parser for it when compared to more complex grammar categories. Given that most
parsers accept context-free grammars, writing fuzzers based on context-free
grammars can be effective in fuzzing the programs that use these parsers.

\section{Method of Evaluation}
\label{sec:evaluation}
In order to get a fair assessment of various fuzzers, it is import to ensure that
we remove as much of external factors as possible that can skew the results. To
ensure that each fuzzer got sufficient time to cache execution paths in memory,
we started with a small warm up loop of \warmuploop iterations. Next, to avoid
skew due to different seeds, we chose random seeds from zero to nine\footnote{
The random seed is used to initialize the pseudorandom number generator. While
the seed values are close to each other, the random numbers generated from the
initial seed values are not close to each other, as even small differences in
bit patterns have a large impact. Hence, we chose to use the numbers from 0 to 9
to be as unbiased as possible (a set of random, random seeds may have an
unforeseen bias).
} and computed the average of \numloops runs using these seeds.

We needed to make sure that there was a level playing ground for all grammar
fuzzers. For grammar fuzzers, it is easier to produce relatively flat inputs
such as say \emph{true} or \emph{false} in JSON grammar than one that requires
multiple levels of nesting. However, when fuzzing, these inputs with complex
structure are usually more useful as these inputs have a higher chance of
producing interesting behavior. A metric such as the number of inputs per
second (as used with mutational fuzzers) unfairly penalizes the grammar
fuzzers that produce inputs with complex structure.
Hence, rather than the number of inputs produced, we opted to simply use the
\emph{throughput}---kilobytes of output produced per second as the metric to
judge the fuzzer performance.

We saw that the maximum depth of recursion had an impact on the throughput.
Hence, we evaluated all the fuzzers (that allowed setting a maximum depth)
with similar depth of recursion, with depth ranging from 8 to 256, with the
timeout set to an hour (36,00 seconds).

Our experiments were conducted on a \emph{Mac OS X} machine with nacOS
\emph{10.14.5.} The hardware was \emph{MacBookPro14,2} with an
\emph{Intel Core i5} two-core processor. It had a speed
of 3.1 Ghz, an L2 cache of 256 KB, and L3 cache of 4 MB. The memory was 16 GB.  All tools, including our own, are single-threaded.  Times measured and reported is the sum of user time and system time spent in the respective process.

\section{Controlling Free Expansion}
\label{sec:depth}
\begin{figure}[H]
\begin{lstlisting}[style=python, numbers=left, numberstyle=\tiny]
def d_key(key, seen):
   if key not in grammar: return 0
   if key in seen: return inf
   return min(d_alt(rule, seen | {key})
          for rule in grammar[key])

def d_alt(rule, seen):
    return max(d_key(s, seen) for s in rule) + 1

\end{lstlisting}
\caption{Computing the minimum depth of expansion $\mu$-depth}
\label{fig:expansiondepth}
\end{figure}
The simple approach given in \Cref{fig:simplefuzzer} is naive in that it provides
no control over how far to go when expanding the \nonterminal tokens. If the grammar
is highly recursive, it can lead to stack exhaustion before all \nonterminal symbols
are effectively expanded to a \terminal string. There is a solution to this problem\footnote{
The text book approach given in
the chapter on Grammar Fuzzing~\cite{fuzzingbook2019:GrammarFuzzer} provides two
controls --- the maximum number of \nonterminal symbols and the number of \emph{trials} to
attempt before giving up.
}. Given a context-free grammar, for any given key, let us define a minimum depth of expansion ($\mu$-depth) as the minimum number of levels of expansion needed (stack depth) to produce
a completely expanded string corresponding to that key. Similarly the $\mu$-depth of a rule is the
maximum of $\mu$-depth required for any of the tokens in the rule.
One can hence compute the $\mu$-depth for each of the \nonterminal symbols.
The idea is similar to the fuzzer in \Cref{fig:simplefuzzer}.
Given a token, check if the token is a \nonterminal. If it is not, the $\mu$-depth is zero.
If it is a \nonterminal, compute the $\mu$-depth of each of the alternative expansion rules
corresponding to it in the grammar. The \nonterminal's $\mu$-depth is the minimum of the $\mu$-depth
of its corresponding rules. If we detect recursion, then the $\mu$-depth is set to $\infty$.
The algorithm for $\mu$-depth computation is given in~\Cref{fig:expansiondepth}.

Once we compute the minimum depth of expansion for every key, one can modify the naive
algorithm in \Cref{fig:simplefuzzer} as follows: Start the generation of input with a
maximum free-stack budget. When the number of stack frames go beyond the budget, choose
only those expansions that have the minimum cost of expansion. With this restriction,
we can now compute the throughput of our fuzzer for the \expr grammar: 103.82~KiB/s with the free expansion depth set to eight, and a maximum of 133~KiB/s when the free expansion depth is set to 32.

\subsection{Precomputed Minimum Depth Expansions}
\label{sec:pool}

We have precomputed the expansion cost. But is our implementation optimal? On
tracing through the program execution, one can see that once the free stack budget is
exhausted, and the program switches to the minimum depth strategy, there is only
a small pool of strings that one can generate from a given key, and all of them have
exactly the same probability of occurrence. Hence, we can precompute this pool.
Essentially, we produce a new grammar with only the minimum depth expansions for
each \nonterminal, and exhaustively generate all strings using an algorithm
similar to \Cref{fig:simplefuzzer}. Precomputing string pools gives us a
throughput of 371.76 kilobytes per second for an expansion depth of 8, and
a throughput of 420.14 for an expansion depth of 32.

Precomputing this pool of strings for each keys, and using the pool when the
free stack budget is exhausted gets us only a modest improvement. Is there any
other optimization avenue?
\begin{figure}
\begin{lstlisting}[style=python, numbers=left, numberstyle=\tiny]
def gen_key(key, depth):
   if key not in grammar: return key
   rules = grammar[key]
           if depth < max_depth else pool[key]
   return gen_alt(random.choice(rules)), depth+1)

def gen_alt(alt, depth):
   return ''.join(gen_key(t, depth) for t in alt)
    
grammar = ...
pool = precompute_pool(grammar)
max_depth = ...
gen_key('<start>', 0)
\end{lstlisting}
\caption{A fuzzer with precomputed string pools}
\label{fig:poolfuzzer}
\end{figure}

\section{Compiling the Grammar}
\label{sec:compiling}

One of the advantages of using precomputed string pools is that it eliminates the cost
of checking whether a token is a \nonterminal or not, and also looking up the
corresponding rules of expansion for the \nonterminal. Can we eliminate this lookup completely?

A grammar is simply a set of (possibly recursive) instructions to produce output.
We have been interpreting the grammar rules using the fuzzer. Similar to how one
can compile a program and generally make it faster by removing the interpreting
overhead, what if we compile the program so that the resulting program produces the fuzzer output?

The idea is to essentially transform the grammar such that each \nonterminal becomes
a function, and each \terminal becomes a call to an output function
(e.g \texttt{print}) with the symbol as argument that outputs the corresponding \terminal symbol. The functions corresponding to \nonterminal symbols have conditional
branches corresponding to each expansion rule for that symbol in the grammar. The
branch taken is chosen randomly. Each branch is a sequence of method calls
corresponding to the \terminal and \nonterminal symbols in the rule.

As before, we incorporate the optimization for \texttt{max_depth} to the functions.
A fragment of such a compiled grammar is given in~\Cref{fig:compiled}.
\begin{figure}[H]
\begin{lstlisting}[style=python, numbers=left, numberstyle=\tiny]
 def expr(depth):
    d = depth + 1
    if d > max_depth: return choice(s_expr)
    c = random.randint(3)
    if c == 0: s = [term(d), "*", term(d)]
    if c == 1: s = [term(d), "/", term(d)]
    if c == 2: s = [term(d)]
    return ''.join(s)
...
\end{lstlisting}
\caption{A fragment of the compiled \texttt{expression} grammar. The \texttt{s_expr}
contains the possible minimum cost completion strings for \texttt{expr} \nonterminal.}
\label{fig:compiled}
\end{figure}
Compiling the grammar to a program, and running the program gives us a faster fuzzer,
with a throughput of 562.22 kilobytes per second for a free expansion depth of 8, and
714.08 kilobytes per second for a free expansion depth of 32.
The pure random Python fuzzer in the Fuzzingbook chapter on
Fuzzing~\cite{fuzzingbook2019:Fuzzer} achieves 1259 kilobytes per second. That
is, this represents a slowdown by a factor of two.

\section{Compiling to a Faster Language}
\label{sec:c}

While Python is useful as a fast prototyping language, it is not
a language known for the speed of programs produced. Since we are
compiling our grammar, we could easily switch the language of
our compiled program to one of the faster languages. For this
experiment we chose \emph{C} as the target. As before, we transform
the grammar into a series of recursive functions in C.

As we are using
a low level language, and chasing ever smaller improvements, a number of things can
make significant difference in the throughput. For example, we no longer do dictionary
lookups for string pools. Instead, we have opted to embed them into the produced
program, and do array lookups instead. Similarly, we use a case statement that directly
maps to the output of the modulus operator. With the new compiled grammar, we reach
a throughput of 14,775.99 kilobytes per second for a depth of 8 and 15,440.12
kilobytes per second for a depth of 32.

\section{The Grammar as a functional DSL}
\label{sec:dsl}

We have seen in the previous section how the grammar can be seen as a domain
specific language for generating strings, and the compiled fuzzer as
representing the grammar directly. Another thing to notice is that the language
of grammar is very much a pure functional language, which means that the ideas
used to make these languages faster can be applied to the compiled grammar. We
examine two such techniques.

\subsection{Partial Evaluation}
\label{sec:pe}

The first
is partial evaluation~\cite{jones1993partial,sorensen1994towards}. The idea of
partial evaluation is simple. In most programs, one does not have to wait until
runtime to compute a significant chunk of the program code. A large chunk can be
evaluated even before the program starts, and the partial results included in
the program binary directly. Indeed, pre-computing the minimum depth expansions
is one such. We can take it further, and eliminate extraneous choices such as
grammar alternatives with a single rule, and inline them into the produced
program. We could also embed the expansions directly into parent expansions,
eliminating subroutine calls. Would such inlining help us?

Using this technique, gets us to 13,945.16 kilobytes per second for a
depth of 8 and 15021.02 kilobytes per second for a depth of 32. That is,
partial evaluation by simple elimination of choices is not very helpful.

\subsection{Supercompilation}
\label{sec:supercompiling}

Another technique is \emph{supercompilation}---a generalization of partial
evaluation~\cite{sorensen1994towards} that can again be adapted to context-free grammars.\footnote{ The original supercompiler is due to Turchin~\cite{turchin1986the},
and involves removing redundancies present in the original program
by abstract interpretation. The language in which it was implemented
\emph{Refal}~\cite{turchin1989refal} is a term rewriting language
reminiscent of the syntax we use for representing the context-free
grammar. }

The idea of supercompilation is as follows: The program generated is
interpreted abstractly, using symbolic values as input (\emph{driving}). During
this abstract interpretation, at any time the interpretation encounters
functions or control structures that does not depend on input, the execution is
inlined. When it encounters variables that depend on the input, a model of the
variable is constructed, and the execution proceeds on the possible
continuations based on that model. If you find that you are working on a
similar expression (in our case, a \nonterminal expansion you have seen before)
\footnote{This is called \emph{renaming} in the language of supercompilation},
then terminate exploring that thread of continuation, produce a function
corresponding to the expression seen, and replace it with a call
to the function \footnote{This is called \emph{folding}.}. During this process,
redundant control and data structures are trimmed out, leaving a residual
program that performs the same function as the original but with a lesser number
of redundant steps.

In the case of the language of context-free grammars, the input is given by the
random stream which is used to determine which rule to expand next. Hence, during the
process of supercompilation, any \nonterminal that has only a single rule gets
inlined.
Further, parts of expansions that have deterministic termination are also inlined,
leaving only recursive portions of the program as named functions. Finally,
the functions themselves are transformed. The \nonterminal symbols are no longer
functions themselves. Instead, they are inlined, and the individual rule
expansions become functions themselves. We will see how these functions lead to a
newer view of the fuzzer in the next sections.

\Cref{fig:supercompiledc} shows a fragment of the supercompiled expression grammar.
\begin{figure}[H]
\begin{lstlisting}[language=C, numbers=left, numberstyle=\tiny]
void expr_1(int depth) {
  int d1 = depth - max_depth;
  if (d1 >= 0) {
    out(s_term[random() % n_s_term]);
  } else {
    int d2 = d1 - 1;
    switch(random() % 5) {
    case 0:
      if (d2 >= 0) {
        out(s_factor[random() % n_s_factor]);
      } else {
        switch(random() % 5) {
        case 0:
          factor_0(depth+4);
...
\end{lstlisting}
\caption{A fragment of the supercompiled grammar in C. As before, \texttt{s_expr} is a
precomputed array of minimum cost completion strings for \nonterminal \texttt{expr}. The function
\texttt{expr_1} represents the second production rule for \texttt{expr}. The function corresponding to the \nonterminal \texttt{expr} itself is inlined}
\label{fig:supercompiledc}
\end{figure}
Supercompiled grammar fuzzer produces a throughput of 14,441.95 kilobytes
at a depth of 8 and 14,903.12 kilobytes at a depth of 32. Supercompilation
does not make much of a difference in the case of \expr, but as we will
see later in the paper, it helps with other subjects.

\section{System-Level Optimizations}
\label{sec:system}

Compilation can improve the performance of interpreted languages, while
supercompilation can improve the performance of functional languages.
We have not yet considered how both \emph{interact with their environment,}
though.  Generally speaking, a fuzzer needs two functionalities: \emph{random
numbers} required to choose between expansions, and \emph{system output} to send
out the produced string. We will see how to optimize them next.

\subsection{Effective Randomness}
\label{sec:random}

Is our generated fuzzer the best one can do? While profiling the fuzzer, two things
stand out. Generating random values to choose from, and writing to a file. Can we
improve these parts? Looking at how random values are used, one can immediately see
an improvement. The pseudo-random generators (PRNG) used by default are focused on
providing a strong random number with a number of useful properties. However, these
are not required simple exploration of input space. We can replace the default PRNG
with a faster PRNG\footnote{We used the \texttt{Xoshiro256**} generator
\url{http://xoshiro.di.unimi.it/xoshiro256starstar.c}}. Second,
we use the modulus operator to map a random number to the limited number of rules
corresponding to a key that we need to choose from. The modulus and division
operators are rather costly. A more performant way to map a larger number to a smaller
number is to divide both and take the upper half in terms of bits.
Another optimization is to recognize that we are rather wasteful in using the random
number generator. Each iteration of PRNG provides us with a 64 bit number, and the
number of choices we have rarely exceed 256. That is, we can simply split the
generated random number to eight parts and use one part at a time. Finally, for
better cache locality, it is better to generate the needed random numbers at one
place, and use them one byte at a time when needed. In fact, we can pre generate
a pool of random numbers, and when the pool is exhausted, trigger allocation of
new random bits, or when the context permits, reuse the old random bits in
different context.
These micro optimizations can provide us with a rather modest improvement.
The throughput is 11,733.98 kilobytes per second for depth of 8, and
22,374.40 kilobytes per second for depth of 32.

\subsection{Effective Output}
\label{sec:output}

As we mentioned previously, output is one of the more performance intensive operations.
Ideally, one would like to generate the entire batch of outputs in memory, and write
to the file system in big chunks.

Switching from \texttt{fputc} to generating complete items, and writing them one
at a time gave us a throughput of 19,538.411 kilobytes per second for a depth of
8 and 69,810.755 kilobytes per second for a depth of 32.

One way to improve the output speed is to use memory mapped files to directly write
the output.
One of the problems here is that for the \texttt{mmap()} call, one need to know the size
of the file in advance. We found that we can tell the operating system to map the file
with the maximum size possible, which the OS obeys irrespective of the actual availability
of space. Next, we can write as much as required to the mapped file. Finally, we call
truncate on the file with the number of bytes we produced. At this point, the operating
system writes back the exact amount of data we produced to the file system.

Unfortunately, \texttt{mmap()} performance was rather variable. We obtained
a throughput of 10,722.41~KiB/s on a depth of~8, and 56,226.329~KiB/s on a depth of~32, and we found this to fluctuate. That is,
depending on \texttt{mmap()} should be considered only after taking into
consideration different environmental factors such as operating system, load on
the disk, and the memory usage.


A fuzzer does not have to write to a file, though.  If the program under test can read input directly from memory, we can also have the fuzzer write only to memory, and then pass the written memory to the program under test.  Obviously, skipping the output part speeds up things considerably; we obtain a throughput of 81,764.75~KiB/s on a depth of~32.  (In the remainder of the paper, we will assume we write to a file.)

\section{Production Machines}
\label{sec:vm}

Is this the maximum limit of optimization? While we are already compiling the grammar to an
executable, our supercompiled program is in a very strong sense reminiscent of a virtual
machine. That is, in our case, the random bits that we interpret as a choice between
different rules to be used to expand can be considered the byte stream to interpret for
a virtual machine. The usual way to implement a virtual machine is to use switched dispatch.
Essentially, we implement a loop with a switch statement within that executes the selected
opcode.

\subsection{Direct Threaded VM}
\label{sec:direct}

However, one of the most performant ways to implement a virtual machine is something called a
\emph{threaded code}~\cite{bell1973threaded,ertl2003the}. A pure bytecode interpreter using
switched dispatch has to fetch the next bytecode to execute, and lookup its definition in
a table. Then, it has to transfer the execution control to the definition. Instead, the idea
of direct threading is that the bytecode is replaced by the starting address of its definition.
Further, each opcode implementation ends with an epilogue to dispatch the next opcode.
The utility of such a technique is that it removes the necessity of the dispatch table, and the central
dispatch loop.
\begin{figure}
\begin{lstlisting}[language=C, numbers=left, numberstyle=\tiny]
produce_member_0: {
    ++returnp;
    *returnp = &&return__0__0__member;
    val = map(2);
    goto *produce_ws[val];
return__0__0__member:;
    *returnp = &&return__1__0__member;
    val = map(1);
    goto *produce_string[val];
return__1__0__member:;
    *returnp = &&return__2__0__member;
    val = map(2);
    goto *produce_ws[val];
return__2__0__member:;
    *out_region++ = ':';
    *returnp = &&return__4__0__member;
    val = map(1);
    goto *produce_element[val];
return__4__0__member:;
    --returnp;
    goto **returnp;
}
\end{lstlisting}
    \caption{A fragment of the threaded interpreter for grammar VM that generates a JSON object. Note the similarity of opcodes to the generated functions from supercompilation.}
    \label{fig:threaded}
\end{figure}
Using this technique gets us to 17,848.876 kilobytes per second for a depth of 8
and 53,593.384 kilobytes for a depth of 32.

\subsection{Context Threaded VM}
\label{sec:context}

One of the problems with direct threading~\cite{ertl2003the} is that it tends to
increase branch misprediction. The main issue is that \texttt{computed goto} targets
are hard to predict. An alternative is context threading~\cite{berndl2005context}.
The idea is to use the \texttt{IP} register consistently, and use \texttt{call} and \texttt{return} based on the value of \texttt{IP} register when possible. This is different
from simply using subroutine calls as no parameters are passed, and no prologue and 
epilogue for subroutine calls are generated. Doing this requires generating assembly, as
\emph{C} does not allow us to directly generate naked \emph{call} and \emph{return}
instructions. Hence, we generated \emph{X86-64} assembly corresponding to a context threaded
VM, and compiled it. A fragment of this virtual machine in pseudo-assembly is given in \Cref{fig:cthreaded}.
\begin{figure}[H]
\begin{lstlisting}[language=python, numbers=left, numberstyle=\tiny]
gen_member_0:
    val = map(2)
    call *gen_ws[val]
    val = map(1)
    call *gen_string[val]
    val = map(2)
    call *gen_ws[val]
    *out_region = ':'
    incr out_region
    val = map(1)
    call *gen_element[val]
    ret
\end{lstlisting}
    \caption{A fragment of the context threaded interpreter for grammar VM that generates a JSON object.}
    \label{fig:cthreaded}
\end{figure}
Context threading got us to 14,805.989 kilobytes per second for a depth of 8 and 16,799.153.
While for \expr, the direct threading approach seems slower than the context threading,
as we will see later, \expr is an outlier in this regard. The context threading approach
generally performs better.
This final variant is actually our fastest fuzzer, which we call the \fone fuzzer.

\section{Evaluation}
\label{sec:comparison}

\begin{figure*}[p]
\begin{subfigure}{\textwidth}
\includegraphics[width=\textwidth]{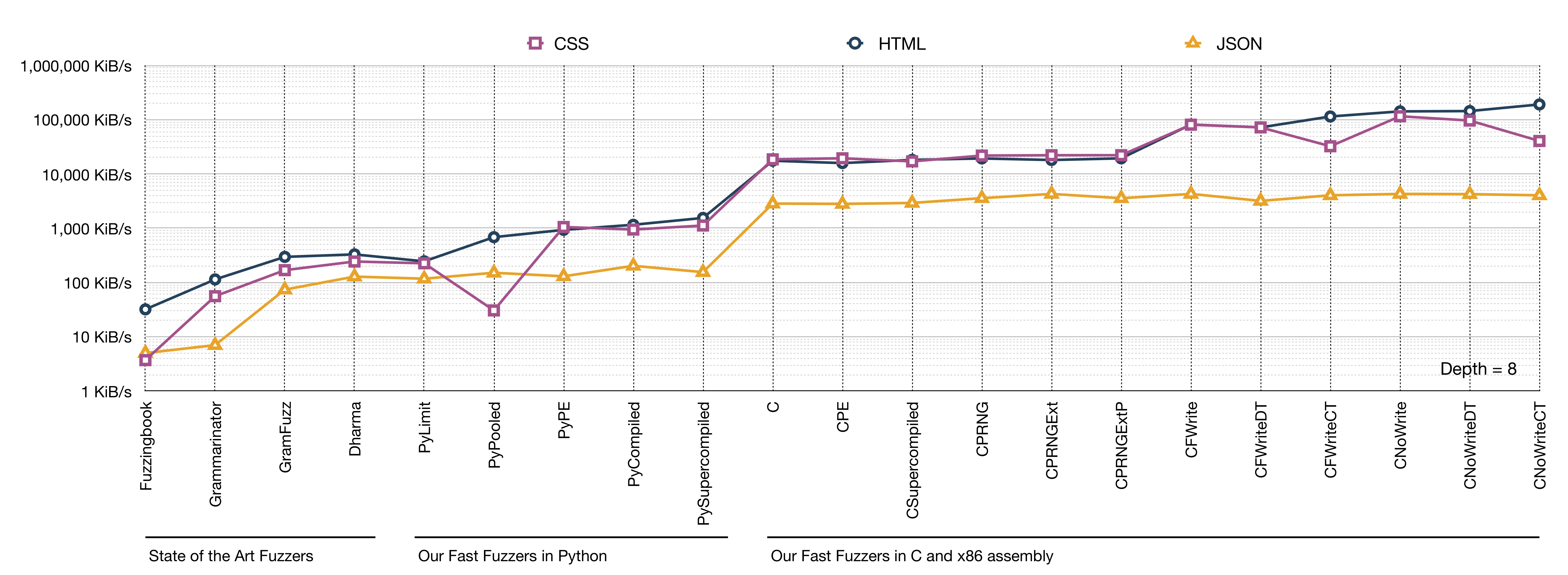}
\end{subfigure}
\begin{subfigure}{\textwidth}
\includegraphics[width=\textwidth]{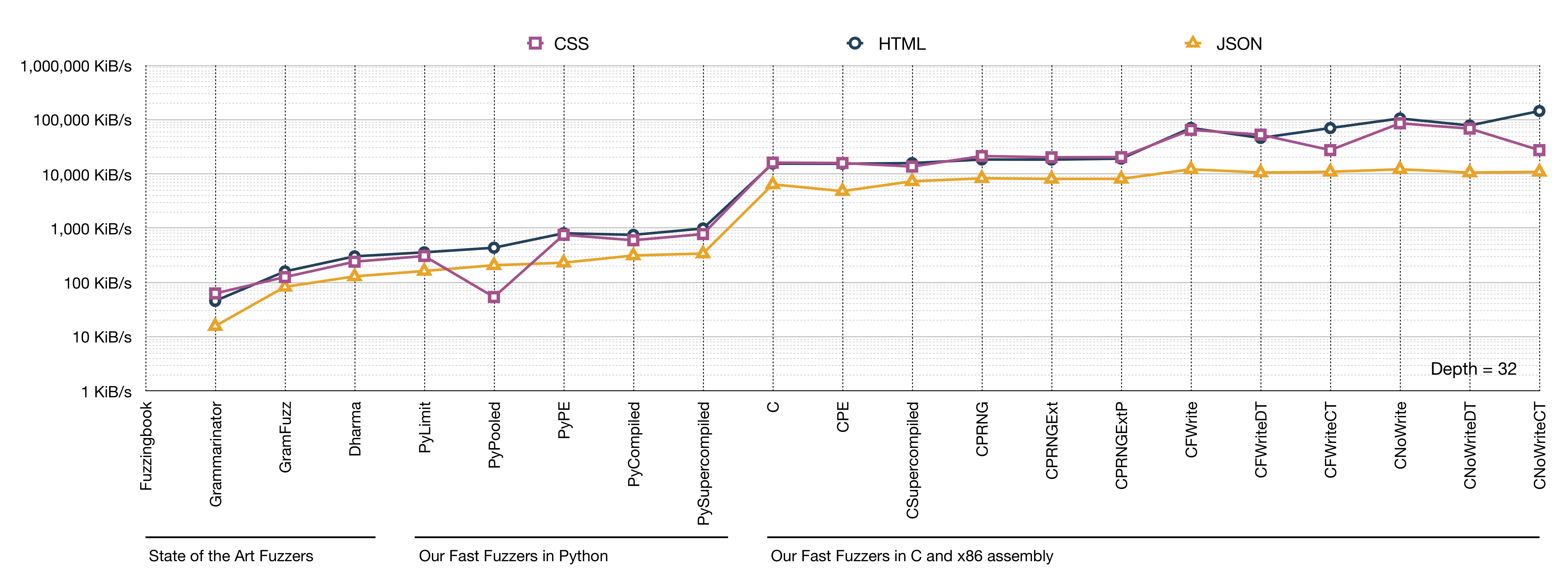}
\end{subfigure}
\begin{subfigure}{\textwidth}
\includegraphics[width=\textwidth]{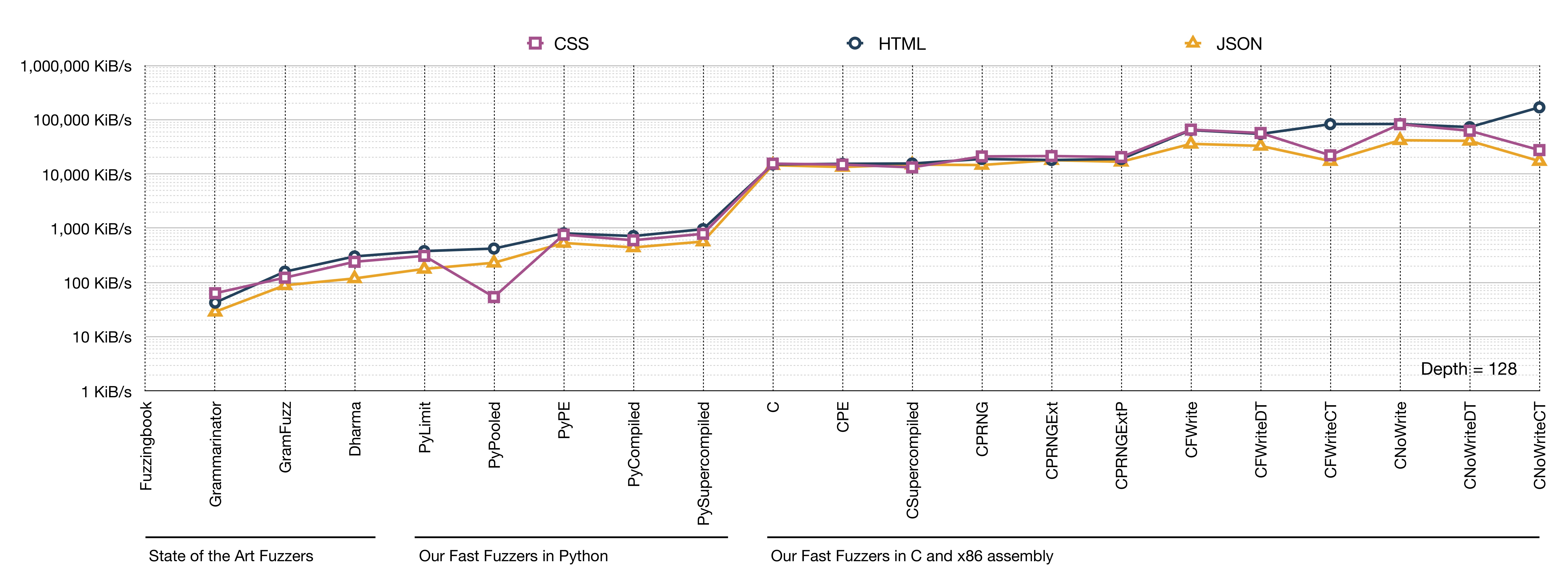}
\end{subfigure}
\caption{Fuzzer throughput with different grammars and depth limits.  
\textbf{PyLimit} = Python with limit (\Cref{sec:depth}).
\textbf{PyPooled} = Python with precomputed string pools (\Cref{sec:pool}).
\textbf{PyPE} = Python with partial evaluation (\Cref{sec:supercompiling}).
\textbf{PyCompiled} = Compiled Python producer (\Cref{sec:compiling}).
\textbf{PySuperCompiled} = Supercompiled Python producer (\Cref{sec:supercompiling}).
\textbf{C} = Compiled C producer (\Cref{sec:c}); with partial evaluation (\textbf{CPE}; \Cref{sec:pe}).
\textbf{CSuperCompiled} = Supercompiled C producer (\Cref{sec:supercompiling}).
\textbf{CPRNG} = C producer with faster random number generator (\Cref{sec:random}) and precompiled random numbers (\textbf{CPRNGExt}).
\textbf{CFWrite} = C Producer Machine (\Cref{sec:vm}) using \texttt{fwrite()} (\Cref{sec:output}); with direct treading (\textbf{CFWriteDT}, \Cref{sec:direct}) and context treading (\textbf{CFWriteCT}, \Cref{sec:context}).
\textbf{CNoWrite, CNoWriteDT, CNoWriteCT} = C Producer Machine writing to memory (\Cref{sec:output}).
}
\label{fig:throughput}
\end{figure*}

So far, we have checked the performance of our fast fuzzing techniques only on one, admittedly very simple grammar.  How do they fare when faced with more complex input formats?  And how do they compare against state-of-the-art grammar fuzzers?  To this end, we evaluate our techniques on three grammars with well-known and nontrivial industry formats (CSS, HTML, and JSON), and compare them against four state of the art tools.

Our results are summarized in \Cref{fig:throughput}, using three different settings for the maximum depth (8, 32, and 128).  The vertical axis lists the throughput achieved by each tool for the respective grammar; note the usage of a logarithmic scale to capture the large differences.  To account for differences due to randomness, the state-of-the-art tools were run 2~times with 1,000 inputs generated per run; our (faster) tools were run 100~times, also with a 1,000 inputs generated per run.  Times listed are averages over all runs and inputs generated.

\subsection{Textbook Fuzzer}

The fuzzingbook from Zeller et al.~\cite{fuzzingbook2019:GrammarFuzzer} specifies a variety of grammar based fuzzers.  We chose
the simplest one with no frills in the interest of performance.  The grammar syntax is somewhat similar to our own but uses strings for rules with specially demarcated tokens for \nonterminal symbols.

The authors in~\cite{fuzzingbook2019:GrammarFuzzer} make it clear that their interest is in teaching how grammar-based fuzzers work, and that performance is not their main goal.  This also shows in our evaluation: The throughput of the fuzzingbook \texttt{GrammarFuzzer} is~3.7 KiB/s for CSS, 31.9~KiB/s for HTML, and 5~KiB/s for JSON.  As it comes to producing high volumes of input, this marks the lowest end of our evaluation; for higher maximum depths than 8, it would not produce results in time at all.

\subsection{Grammarinator}

\grammarinator~\cite{hodovan2018grammarinator} is the state of the art fuzzer by Ren\'ata Hodov\'an et al.
It is written in Python and accepts a context-free grammar in the \emph{ANTLR} grammar format.
The \grammarinator is the only fuzzer in our set of competing fuzzers that compiles the grammar
to Python code first before fuzzing. \grammarinator is also the only grammar fuzzer included in the
\emph{BlackArch}\footnote{\url{https://www.blackarch.org/}} Linux distribution for pen testers.

\grammarinator is faster than the fuzzingbook grammar fuzzer: Given a maximum depth of 8, it achieves a throughput of 55.8~KiB/s, 113.7~KiB/s, and 7~KiB/s for CSS, HTML, and JSON grammars respectively.  It is slower, however, than the other state-of-the-art tools, not to speak of the fast fuzzers introduced in this paper.  This may be due to \grammarinator spending some effort in \emph{balancing} its outputs.  \grammarinator maintains a \emph{probability} for each expansion alternative, and every time an expansion is chosen, it reduces this probability, thus favoring other alternatives.  This balancing costs time, and this slows down \grammarinator.  The benefit could be a higher variance of produced strings, but this is a feature we did not evaluate in the present paper.

\subsection{Gramfuzz}

\gramfuzz~\cite{guo2013gramfuzz} is another grammar based fuzzer written in
Python.\footnote{Yes, all grammar fuzzers we know of are written in Python.  Python seems to be the language of choice for grammar fuzzers.} It uses its own format for specifying a context-free grammar. The
fuzzer allows specifying a recursion depth to limit expansions.

Due to its simplicity, \gramfuzz wins over \grammarinator for all grammars in our evaluation.  Again, given a maximum recursion depth of 8, it achieves a throughput of 168.3~KiB/s, 295.6~KiB/s, and 73.8~KiB/s for CSS, HTML, and JSON grammars respectively, which is 5--10 times as fast as \grammarinator.  This is the more interesting as \gramfuzz interprets the grammar structure instead of compiling it to code.

\subsection{Dharma}

The \dharma\footnote{\url{https://github.com/MozillaSecurity/dharma}} fuzzer
from Mozilla~\cite{mozilla2019dharma} is a grammar based fuzzer that has
received wide attention from the industry. It is (again) written in Python, and
uses its own custom format for specifying the grammar. We found that the
\dharma grammar syntax was a little unwieldy in that there is no way to specify
empty rules, and empty strings. Hence, we had to work around this using white
space for our evaluation.

\dharma's throughput again improves over \gramfuzz; with a recursion depth of 8, it reaches 242~KiB/s, 328.9~KiB/s, and 128.1~KiB/s for CSS, HTML, and JSON grammars respectively.  This makes \dharma the fastest state-of-the-art fuzzer, and hence the baseline our faster techniques can compare against.

\subsection{Fast Python Fuzzers}

We now discuss our own fast fuzzers written in Python and/or compiling to Python code.  \textbf{PyLimit} is the very simple fuzzer with depth control introduced in \Cref{sec:depth}.  With 225.3~KiB/s, 244.8~KiB/s, and 117.2~KiB/s for CSS, HTML, and JSON grammars respectively, it is only marginally slower than \dharma.  This is the baseline our other fuzzers compare against.  We see that partial evaluation (\textbf{PyPE}, 1051.3~KiB/s, 931.2~KiB/s, 129.7~KiB/s) already brings large speedups.  Supercompilation (\textbf{PySuperCompiled}) then boosts the throughput to 1,119.2~KiB/s, 1,544.5~KiB/s, and 154.9~KiB/s.  For JSON, however, supercompilation is slower than regular compilation (202.7~KiB/s); such effects can occur if the generated code is too large and hits processor caching limits.

\subsection{Fast Fuzzers in C and x86 assembly}

Switching from Python to C as the language for producers and compiling the C code to machine code brings a huge performance gain.  For a maximum depth of 8, the throughput of the compiled C producer (\textbf{C}) is 18,558.9~KiB/s, 17,481.3~KiB/s, and 2,838.7 KiB/s on CSS, HTML, and JSON grammars respectively.  This is about 20~times as fast as the compiled Python producer (\textbf{PyCompiled}).  Partial evaluation and supercompilation bring further benefits, notably with higher maximum recursion depths (32~and~128).  Using pseudo-random number generators speed up producers by about 25\%, as do precompiled random numbers.

The next big performance boost comes from introducing producer machines.  Using a producer machine (\textbf{CFWrite}) that also incorporates the above random number optimizations, achieves a throughput of 80,722.1~KiB/s, 81,323.7~KiB/s, and 4,281.5~KiB/s, respectively, which again is a factor of four higher than the fastest C producer.  Adding direct and context treading can further improve performance, depending on depth and grammar; on HTML, the \textbf{CFWriteCT} variant achieves a throughput of 141,927.6~KiB/s, which sets a record in our evaluation.  (Actually, we still can be faster, but only by omitting the writing part; the last three columns in our charts (\textbf{NoWrite}) show the throughput when writing to memory only.)


\subsection{Discussion}

All in all, we have seen that our techniques---notably compilation, using C/assembly as producer languages, and finally building dedicated virtual machines---can considerably speed up fuzzing.  Compared to \dharma, the fastest fuzzer in our evaluation, one can expect speedup factors of~100--300.

Does this mean that one can now fuzz several hundred times faster than before?  If we fuzz fast programs, individual functions, or hardware components, we may indeed see significant overall speed ups.  If most of the overall runtime is spent on running the program under test, though, a faster fuzzer will still save CPU time, but the relative benefits of a faster fuzzer will be smaller.  However, the increased speed allows to stress test such programs with much \emph{larger} inputs than before, turning generation time from minutes to tenths of seconds.

Finally, let us not forget that the main ingredient for the speed of grammar-based fuzzing is not so much the optimizations as described in this paper, but the \emph{grammar itself.}  If we were set to compare against random character-based fuzzers such as AFL~\cite{nossum2016filesystem} and counted the number of \emph{valid inputs} produced per second, then even the simplest grammar fuzzer would easily outperform its competition.  This highlights the need for techniques to infer grammars from existing programs~\cite{hoschele2016mining,bastani2017synthesizing,blazytko2019usenix}, as they would end up in an extremely fast and extremely effective fuzzer.

\section{Related Work}
\label{sec:related}

Our related work falls into the following categories.

\begin{description}
\item[Grammars.]
While informally specified grammars have been around from the origins of human language, the first
formalization occurred in 350 BC by Dak{\d{s}}iputra
P{\=a}{\d{n}}ini in Ash{\d{t}}{\=a}dhy{\=a}y{\=\i}~\cite{panini0ashtadhyayi}. Chomsky~\cite{chomsky1956three} introduced the formal models for describing languages, namely:
finite state automatons, context-free grammars, context-sensitive grammars, and universal grammars
in the increasing order of power (called the Chomsky hierarchy).

\item[Model based Fuzzers.]
Generating inputs using grammars
were initially explored by Burkhadt~\cite{burkhardt1967generating}, and later by
Hanford~\cite{hanford1970automatic} and Purdom~\cite{purdom1972a}. Modern fuzzing tools that use
some input models include CSmith~\cite{yang2011finding}, LangFuzz~\cite{holler2012fuzzing},
Grammarinator~\cite{hodovan2018grammarinator} (Python), and Domato~\cite{gratric2019domato} (Python),
Skyfire~\cite{wang2017skyfire} (Python), and Superion~\cite{wang2019superion}, which extends AFL.

\item[Grammar Learning.]
There is a large amount of research on inferring \emph{regular grammars} from
blackbox systems~\cite{higuera2010grammatical,walkinshaw2008evaluation,walkinshaw2007reverse}. The notable algorithms include L*~\cite{angluin1987learning} and RPNI~\cite{oncina1992inferring}.

Blackbox approaches can also be used to learn context-free grammars. Notable approaches
include version spaces~\cite{vanlehn1987a}, and GRIDS~\cite{langley2000learning}.
GLADE~\cite{bastani2017synthesizing} and later REINAM~\cite{wu2019reinam} derives the context-free input grammar focusing on blackbox \emph{programs}.
Other notable works include the approach by Lin et al.~\cite{lin2008deriving} which extracts the AST from programs that parse their input, AUTOGRAM by H{\"o}schele et al.~\cite{hoschele2016mining,hoschele2017active} which learns the input grammar
through active learning using source code of the program, Tupni~\cite{cui2008tupni} by 
Cui et al. which reverse engineers input formats using taint tracking,
Prospex~\cite{comparetti2009prospex} from Comparetti et al. which is able to reverse engineer
network protocols, and Polyglot~\cite{caballero2007polyglot} by Caballero et al.

Another approach for model learning is through machine learning techniques where the model is not
formally represented as a grammar. Pulsar~\cite{gascon2015pulsar} infers a Markov model for representing
protocols. Godefroid et al.~\cite{godefroid2017learn} uses the learned language model to fuzz.
IUST-DeepFuzz from Nasrabadi et al.~\cite{nasrabadi2018neural} uses infers a neural language model
using RNNs from the given corpus of data, which is used for fuzzing.

\item[Faster execution.]
One of the major concerns of fuzzers is the speed of execution --- to be effective,
a fuzzer needs to generate a plausible input, and execute the program under fuzzing.
Given that programs often have instrumentation enabled for tracing coverage, it becomes
useful to reduce the overhead due to instrumentation. The \emph{untracer} from
Nagy et al.~\cite{nagy2019full} can remove the overhead of tracing from parts of the
program already covered, and hence make the overall program execution faster. Another
approach by Hsu et al.~\cite{hsu2018instrim} shows that it is possible to reduce the
overhead of instrumentation even further by instrumenting only the parts that are required
to differentiate paths.

\item[Grammar fuzzers.]
A number of grammar based fuzzers exist that take in some form of a grammar.
The fuzzers such as
Gramfuzz\cite{guo2013gramfuzz}, Grammarinator~\cite{hodovan2018grammarinator},
Dharma~\cite{mozilla2019dharma}, Domato~\cite{gratric2019domato}, and CSS
Fuzz~\cite{ruderman2007introducing}
allow context-free grammars to be specified externally. Other fuzzers~\cite{cui2014novel,wang2017automatic} allow specifying a regular grammar or
equivalent as the input grammar. Some~\cite{dewey2014language} also allow
constraint languages to specify context sensitive features. Other notable
research on grammar based fuzzers include
Nautilus~\cite{aschermann2019nautilus}, Blendfuzz~\cite{yan2013structurized},
and the Godefroid's grammar based whitebox fuzzing~\cite{godefroid2008grammar}.

\item[Optimizations in Functional Languages.]
We have discussed how the fuzzer can be seen as a limited functional
domain specific language (DSL) for interpreting context-free grammars.
Supercompilation is not the only method for optimizing functional programs.
Other methods include deforestation~\cite{wadler1988deforestation}, and
partial evaluation et al.~\cite{jones1993partial}.
Further details on how partial evaluation, deforestation and
supercompilation fit togetherr can be found in
Sorensen et al.~\cite{sorensen1994towards}. 

\item[Optimizations in Virtual Machine Interpreters.]
A number of dispatch techniques exist for virtual machine interpreters. The most
basic one is called switch dispatch in which an interpreter fetches and executes
an instruction in a loop~\cite{brunthaler2009virtual}. In direct threading,
addresses are arranged in an execution thread, and the subroutine follows the
execution thread by using computed jump instructions to jump directly to the
subroutines rather than iterating over a loop. A problem with the direct
threading approach is that it is penalized by the CPU branch predictor as
the CPU is unable to predict where a computed jump will transfer control to.
An alternative is context threading~\cite{berndl2005context} where simple
\emph{call} and \emph{return} instructions are used for transferring control
back and forth from subroutines. Since the CPU can predict where a return would
transfer control to, after a call, the penalty of branch misprediction is
lessened.
\end{description}

\section{Limitations and Future Work}
\label{sec:limitations}

Despite our improvements in speed, our work has lots of room for improvement, notably in terms of supported language features and algorithmic guidance.

What we have presented is a deliberately simple approach to building
grammar based fuzzers. To make the exposition simple, we have chosen to limit
the bells and whistles of our fuzzers to a minimum---just a way to control
the maximum amount of recursion. This allows us to show how to view the
fuzzer first as an interpreter for a programming language, and next as an
interpreter for random bitstream. However, this means that we have left
unimplemented a number of things that are important for an industry grade
fuzzer. For example, the fuzzer does not have a way to use different
probabilities for production rule expansions. Nor does it have a way to accept
feedback from the program under test---for instance, to guide production towards input features that might increase coverage. Similarly, there is no way for it to actually run the program, or to
identify when the program has crashed. Other limitations include the inability
to use a pre-existing corpus of sample data, or to infer input models from such
data.  All these are parts that need to be incorporated to take the fuzzer from an
academic exercise to a fuzzer fully equipped to fuzz real programs. We take
a look at how some of these features might be implemented next.

\subsection{Controlling the Fuzzer}
\label{sec:control}

\begin{figure}[H]
\begin{lstlisting}[style=python, numbers=left, numberstyle=\tiny]
def unroll_key(grammar, key='<start>'):
    return {tuple(rn) for ro in grammar[key]
            for rn in unroll_rule(grammar, ro)}

def unroll_rule(grammar, rule):
    rules = [grammar[key] if key in grammar
                          else [[key]]
            for key in rule]
    return [sum(l, []) for l in product(*rules)]

def unroll(grammar):
    return {k:unroll_key(grammar, k)
            for k in grammar}\end{lstlisting}
    \caption{Unrolling the grammar one level}
    \label{fig:unrolled}
\end{figure}

\begin{figure}[H]
\lstset{
 morekeywords={start,object,array,string,number,cvalue,svalue,l1,l2,l3},
 moredelim={[is][stringstyle]{'}{'}},
}
\begin{lstlisting}[style=python, numbers=left, numberstyle=\tiny]
{
'<start>':  (['<l1>'])
'<l1>':     (['<cvalue>'], ['<l2>'])
'<l2>':     (['<cvalue>'], ['<l3>'])
'<l3>':     (['<cvalue>'], '<svalue>')
'<cvalue>': (['<object>'],
             ['<array>'],
             ['<string>'],
             ['<number>']),
'<svalue>': (["true"], ["false"], ["null"]),\end{lstlisting}
    \caption{Managing the probability of expansion for simple values such as \texttt{true},
    \texttt{false} and \texttt{null} with multi-level embedding. At each level, the probability
    of selection is halved.}
    \label{fig:embed}
    \vspace{-\baselineskip}
\end{figure}

One of the problems with the our simple approach is that some of the grammars
contain rules that are unbalanced, when compared to other rules.
For example,
For example, the top level of a JSON grammar has \texttt{true},
\texttt{false}, and \texttt{null} as top level expansion rules. Since we
randomly choose an expansion rule, these get produced in high frequency, which
is undesirable.

\subsection{Managing probability of strings produced}

One way to prevent this is to unroll the grammar such that the top level
productions with limited expansions are reduced in probability of selection.
\Cref{fig:unrolled} shows how each rule in a grammar can be unrolled by one
level. This can be as many number of times as required to obtain a flatter
grammar.  Another technique for reducing the probability of choice for these
expansions is to embed them in lower levels as shown in \Cref{fig:embed}.
Of course, if a fine control over the probability of choice for each expansion
is desired, one may modify the fuzzer to accept probability annotated grammar
instead.

\subsection{Fine grained fuzzer control}

In \Cref{sec:random} we saw how to pre-allocate random numbers, and use this stream as an input
to the fuzzer. This stream of random numbers also provides us with a means of controlling the
fuzzer. From \Cref{sec:vm}, we have seen how the random numbers are essentially contextual
opcodes for the fuzzing virtual machine. This means that we can specify the path-prefix to be
taken by the fuzzer directly. It necessitates relatively minor change in the output generation
loop in the body of the \texttt{main()}, and not in the core fuzzer.

\subsection{Full supercompilation on the virtual machines}

We have detailed how supercompilation can be used to improve the fuzzer.
However, due to time constraints, and complexity of implementation, we have not
implemented the full supercompilation on the direct and context threading
virtual machines. This will be part of our future work.
 
\subsection{Superoptimization}
\label{sec:superopt}

We note that the assembly program we generated is not optimized as we are not
experts in \emph{x86-64} assembly, and likely has further avenues for
optimization if we had more expertise in house. However, the situation is not
bleak.  There is a technique called
\emph{superoptimization}~\cite{schkufza2013stochastic} that can generate
optimized versions of short sequences of assembly instructions that are loop
free. If you remember from \Cref{sec:supercompiling}, we used
\emph{supercompiling} to generate long sequences that are loop free that
correspond to each opcode we have. Hence, our generated assembly program is
particularly well suited to such optimization techniques. This will be taken
up for our future work.
%

\section{Conclusion}
\label{sec:conclusion}

Fuzzing is one of the key tools in a security professional's arsenal, and
grammar based fuzzers can deliver the best bang for the buck. Recent
developments in grammar inference and faster execution of instrumented programs
puts the focus on improving the speed of grammar based fuzzing.

We demonstrated how one can go from a simple grammar fuzzer in Python with a
low throughput to a grammar fuzzer that is a few orders of magnitude faster,
and near the limit represented by pure random fuzzers. Our fast fuzzer is able
to generate a throughput of more than a hundred megabytes per second, producing
valid inputs much faster than any of the competitors, and even faster than the simplest fuzzer which simply concatenates random characters.

But the pure speed is not our main contribution.  We show that by treating a
context-free grammar as a pure functional programming language, one can apply
approaches from implementation of the functional programming languages domain to
make faster fuzzers.  Next, we show that by treating the random stream as a
stream of opcodes for a fuzzing virtual machine, we can derive more mileage out
of the research on efficiently implementing virtual machine interpreters.  All in all, we have hardly exhausted the possibilities in these spaces, and we look forward to great and fast fuzzers in the future.

We are committed to making our research reproducible and reusable.  All of the source code and data referred to in this paper is available as self-contained Jupyter Notebooks, allowing to run the \fone prototype and fully replicate and extend all experiments.  Visit our repository at

\begin{center}
\url{https://github.com/vrthra/f1}
\end{center}

\printbibliography
\end{multicols}
\end{document}